\PassOptionsToPackage{numbers,sort&compress}{natbib}
\documentclass[10pt,twocolumn,letterpaper]{article}

 \usepackage{cvpr}              

\usepackage[dvipsnames]{xcolor}

\definecolor{cvprblue}{rgb}{0.21,0.49,0.74}
\usepackage[pagebackref,breaklinks,colorlinks,citecolor=cvprblue]{hyperref}

\def\paperID{*****} 
\def\confName{CVPR}
\def\confYear{2024}

\title{Detecting Visual Triggers in Cannabis Imagery: A CLIP-Based Multi-Labeling Framework with Local-Global Aggregation}

\author{Linqi Lu, Xianshi Yu, Akhil Perumal Reddy\\
University of Wisconsin-Madison\\
{\tt\small [llu84, xyu384, perumalreddy]@wisc.edu}
}

\usepackage{amsthm}

\usepackage{graphicx}
\usepackage{tabularray}
\usepackage{color}
\usepackage{comment}

\definecolor{myblue}{RGB}{28,73,144}
\definecolor{mygreen}{RGB}{44,85,17}
\definecolor{myred}{RGB}{239,64,64}

\begin{document}
\maketitle
\begin{abstract}
This study investigates the interplay of visual and textual features in online discussions about cannabis edibles and their impact on user engagement. Leveraging the CLIP model, we analyzed 42,743 images from Facebook (March 1 to August 31, 2021), with a focus on detecting food-related visuals and examining the influence of image attributes such as colorfulness and brightness on user interaction. For textual analysis, we utilized the BART model as a denoising autoencoder to classify ten topics derived from structural topic modeling, exploring their relationship with user engagement. Linear regression analysis identified significant positive correlations between food-related visuals (e.g., fruit, candy, and bakery) and user engagement scores, as well as between engagement and text topics such as cannabis legalization. In contrast, negative associations were observed with image colorfulness and certain textual themes. These findings offer actionable insights for policymakers and regulatory bodies in designing warning labels and marketing regulations to address potential risks associated with recreational cannabis edibles.
\end{abstract}    

\section{Introduction}
\label{sec:intro}

Cannabis edibles are now legally available in certain regions; however, their rising prevalence has raised significant concerns regarding their impact on both public health and the online environment. The consumption of edible cannabis products has been associated with a spectrum of adverse health effects, including cognitive impairment, elevated heart rate, heightened anxiety, and addiction risks \cite{welty2019substance,shi2019impacts}. Furthermore, the visual representation of cannabis-related imagery—characterized by vibrant colors, candy-like appearances, and bakery-inspired themes—has shown a propensity to evoke specific emotional and perceptual responses \cite{sharma2023visual}. Aside from being aesthetic, visual triggers might influence online expressions and shape societal attitudes toward cannabis products, especially among young social media users.     By normalizing cannabis use through online marketing using visual elements like foodie cues, younger audiences may become desensitized to the risks of marijuana use. Given these dynamics, it is crucial to systematically investigate the interplay between visual elements in edible cannabis-related online content and user engagement. Such insights are invaluable for guiding governmental bodies and public health organizations in crafting effective warning labels and regulating marketing strategies to mitigate risks associated with cannabis overuse. Building upon this context, we posed the following research questions:\\
RQ1: How effective is the CLIP model in detecting food-related visual elements (e.g., candy, bakery, and fruit) in social media images, and what is its performance after fine-tuning improvement?\\
RQ2: What is the relationship between food-related visual features and user engagement? \\
RQ3: How do low-level image attributes, specifically colorfulness and brightness, correlate with user engagement on social media platforms? \\
RQ4: How are textual topic frames, such as discussions about marijuana legalization, associated with user engagement in social media posts?

\section{Methods}
\label{sec:data}
\subsection{Data Collection and Processing}
The dataset was extracted from Facebook through the utilization of the Crowd Tangle API. The data is from March 1, 2021, to August 31, 2021. A pre-established list of search keywords, which encompassed terms such as "cannabis," "edibles," "kush," and "marijuana," was deployed to identify posts related to recreational edible marijuana. Then, the image-level collection was based on the post-level information, which resulted in approximately 20,000 images per month. For data filtering, we filtered out noise by searching the keyword through the account name and account description to keep the cannabis-related marketing posts. To enhance the precision of our collected posts and ensure image relevance, we implemented a blacklist to exclude irrelevant content. This blacklist contained terms generally associated with non-edible cannabis consumption, such as 'blunt' and 'dab,' as well as phrases primarily related to smoking, vaping, or topical applications, like 'burning one down' and 'Enail/e-nail.' Additionally, it included mentions of 'hemp' in conjunction with 'flax,' 'chia,' or 'cereal,' to further refine our dataset.
\subsection{Engagement Score}
In CrowdTangle API, user engagement is quantified using a scoring system, where a post's actual interactions are compared to a calculated benchmark based on similar past posts. Scores above 1.0x signify overperformance, while scores below -1.0x indicate underperformance. This system adjusts for posts significantly underperforming their benchmarks, assigning negative scores to emphasize limited engagement. Additionally, posts marginally over or under the expected interaction threshold receive scores between 0x and 1x, providing a refined measure of engagement. This methodology ensures a straightforward yet detailed assessment of user interaction in an academic context.
\subsection{Data Summaries}
There are 42,743 images in the dataset after data cleaning and removing duplicates.\\
\textbf{Data Composition:}
Approximately 95\% of the dataset is comprised of link or photo data, suggesting a diverse range of media content.\\
\textbf{Image Text:}
A significant observation is that approximately 75\% of the image text is null, indicating that a substantial portion of the images lacks textual information.\\
\textbf{Message Length Distribution:}
Analysis of message length reveals that nearly 80\% of the messages contain fewer than 100 words. This suggests that a majority of the content is concise and to the point.\\
\textbf{Temporal Distribution:}
Both the diurnal cycle and weekday distribution exhibit a balanced pattern. This implies that the dataset is evenly distributed across different times of the day and days of the week.
\begin{figure}
  \includegraphics[width=0.9\linewidth]{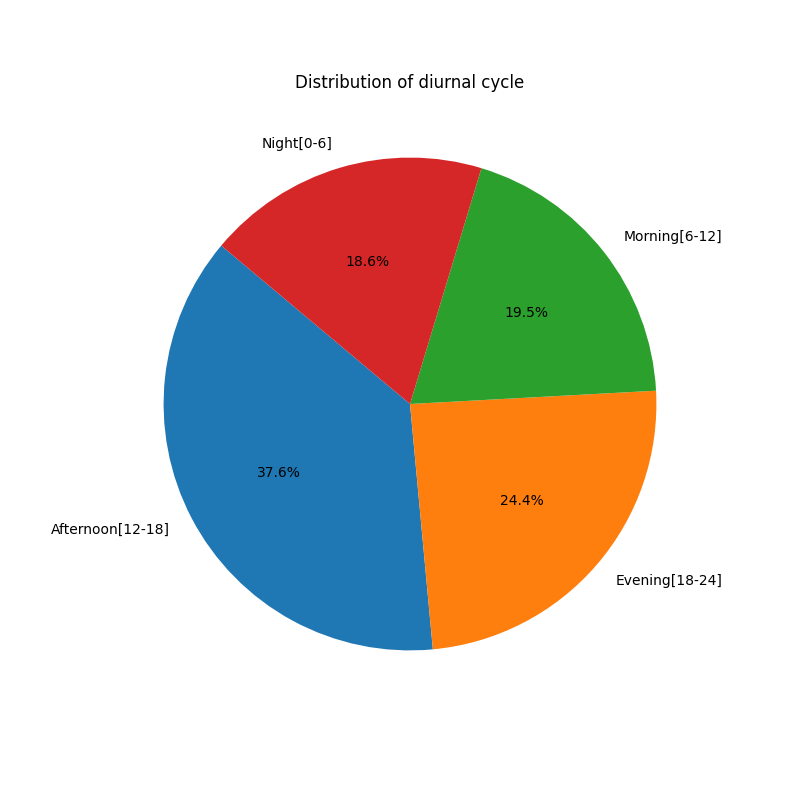}
  \caption{}
  \label{fig:ds1}
\end{figure}

\begin{figure}
  \includegraphics[width=0.9\linewidth]{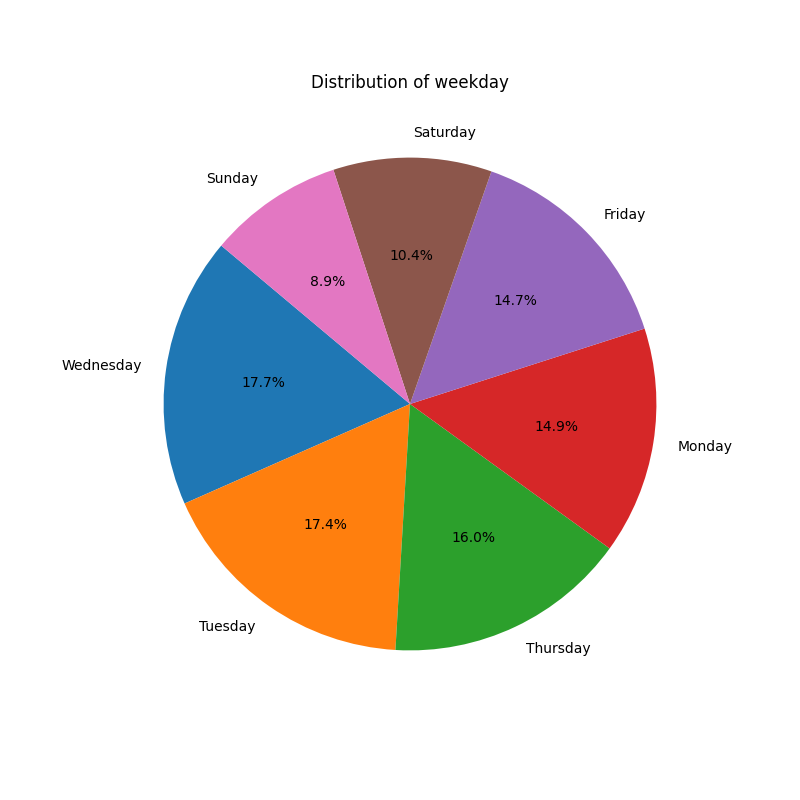}
  \caption{}
  \label{fig:ds2}
\end{figure}

\section{Extract Features From Images and Texts}

\subsection{Colorfulness and Brightness}

The image colorfulness function analyzes the colorfulness of an input image. It achieves this by splitting the image into its Red (R), Green (G), and Blue (B) components. Subsequently, it calculates the differences between the Red and Green channels (rg) and the difference between 0.5 times the sum of Red and Green and the Blue channel (yb). The function then computes the mean and standard deviation for both rg and yb. By combining these statistics using a specified formula, it derives a "colorfulness" metric for the image, emphasizing both mean and standard deviation components \cite{hasler2003measuring}. The final result is returned as a measure of the image's colorfulness.

The brightness function is designed to evaluate the average brightness of an input image. It begins by converting the image to the HSV color space. From the resulting HSV image, the function extracts the Value channel, which represents the brightness information. The average brightness is then calculated by computing the mean of the values in the extracted Value channel. This function provides a quantitative measure of the overall brightness of the input image, useful for understanding the luminance characteristics of the image in terms of its HSV representation.

\subsection{Visual Object Detection}
\label{sec:clip}

We utilized a pre-trained visual language model, Contrastive Language-Image Pre-Training (CLIP) \cite{radford2021learning}, to automatically detect visual elements in each image. The selection of these visual elements was guided by our research questions and insights from existing literature \cite{tan2022presence}. The identified visual elements include {\it candy, bakery, fruit, human, weed}, and {\it text}. This task was formulated as a multi-label classification problem, as each image could contain zero, one, or multiple visual elements (e.g., an image could simultaneously include both {\it human} and {\it candy}).

CLIP was chosen for this task due to its ability to support user-defined labels, a feature not commonly available in other pre-trained image classification or object detection models. CLIP assigns labels by calculating embeddings for both pre-specified text queries and the image itself. The cosine similarity between these embeddings is then computed for each query-image pair. These similarity scores are subsequently used to infer image labels. For instance, probabilities for each label can be derived using a softmax activation function applied to the similarity scores for a given image. Detailed information on the specific queries used is provided in Appendix \ref{app:query}. 

To enhance the performance of CLIP in multi-label classification, we employed the Local-Global Aggregation (LGA) procedure proposed by \cite{abdelfattah2023cdul}.  This method extends the standard CLIP approach by incorporating similarity scores computed for local patches of each image in addition to the overall image-level score. The local and global scores are then aggregated to generate a final similarity score for each query-image pair, referred to as the LGA score.

Thresholding was applied to the LGA scores to assign labels to each image. Thresholds were determined individually for each of the six visual elements by referencing manual annotations created for a randomly sampled subset of images (details provided below). For each element, the threshold maximizing the F1 score for binary classification was selected. The distribution of visual elements after applying these thresholds is summarized in Table \ref{tab:percent}. The results indicate that certain elements, such as {\it human} and {\it text}, are substantially more prevalent than others, such as {\it candy} and {\it fruit}, in our dataset.

\begin{table}
  \centering
  \begin{tabular}{@{}lc|c|c|c|c|c}
    \toprule
     & candy & bakery & fruit & human & weed & text\\
    \midrule
    \% & 0.5 & 1.6 & 1.41 & 40.8 & 15.0 & 58.8 \\
    \bottomrule
  \end{tabular}
  \caption{Percentage of images labeled with each visual element.}
  \label{tab:percent}
\end{table}

\subsection{Performance of CLIP Multi-Labeling}

\textbf{Performance Evaluation.} Before integrating the CLIP-generated image labels into downstream analyses, we evaluated their accuracy by comparing them to manual annotations from a randomly sampled subset of images. The evaluation revealed a critical limitation: the absence of explicit negative classes in the token list led to frequent misclassification. For instance, when labeling an image of a pet using tokens such as `weed,’ `text,’ `candy,’ `bakery,’ `fruit,’ `human,’ and `others,’ the model often misclassified the image as `human,’ assigning it a high probability score. However, when the token `pet’ was added to the list, the model no longer misclassified the image as `human.’ This result highlighted the importance of including tokens that capture unrelated or ambiguous categories to improve the robustness of CLIP's multi-labeling process.

\textbf{Token Refinement Approach.} To address the identified limitations, we applied an image captioning model to all collected images, extracted common topics from the generated captions, and incorporated irrelevant or ambiguous topics into the token list for the `others’ category. This refinement was intended to mitigate misclassification and improve the model’s ability to identify relevant visual cues.

\textbf{Manual Annotation for Validation.} We manually annotated a random sample of 1,049 images to validate the CLIP-generated labels. Three undergraduate annotators labeled each image following a detailed codebook, achieving high inter-coder reliability (Krippendorf's $\alpha > 0.85$). The manually annotated dataset served as a benchmark to evaluate the performance of the CLIP model and guided refinements to the labeling process.

\textbf{Fine-Tuning for Improvement.} Based on the manual annotations, we identified opportunities to improve the model's performance. To achieve this, we fine-tuned the pre-trained CLIP model using the manually annotated dataset. 

\begin{figure}
  \includegraphics[width=0.9\linewidth]{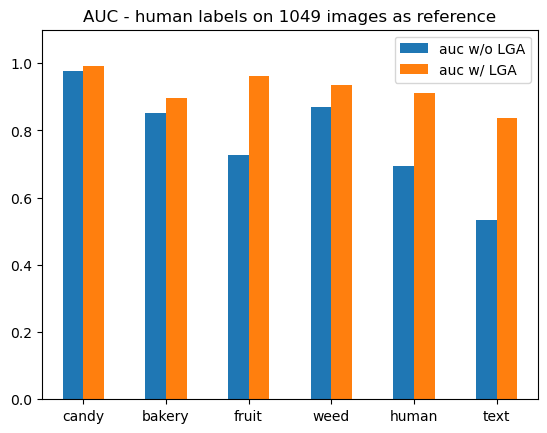}
  \caption{Evaluation of CLIP model performance in image labeling. The figure presents the Area Under the Curve (AUC) scores, using manual annotations as the reference standard. The results compare the performance of the CLIP model with Local-Global Aggregation (LGA) scores (orange) against the baseline performance of the CLIP model without LGA (blue), where similarity scores for the entire image were thresholded directly.}
  \label{fig:auc}
\end{figure}

{\bf Performance of CLIP multi-labeling.} Before using the CLIP-generated image labels in our preliminary analysis, we examined the performance of the LGA scores by comparing them with manual labels of the 1049 images. Figure \ref{fig:auc} shows that, the LGA scores performs reasonably well in multi-labeling our images. This is demonstrated by 0.9 plus AUC (Area under the ROC Curve) for almost all visual elements, with human manual labels as reference. The results in Figure \ref{fig:auc} also shows that the LGA procedure is necessary when performing multi-labeling on our data, as demonstrated by the large drop of AUC when the LGA procedure is not used.

\subsection{Identify Topics in Post Texts}

We utilized the pre-trained Bidirectional and Auto-Regressive Transformers (BART) \cite{lewis2020bart} model to extract topics from the post texts, where ten topics were selected using Structural Topic Modeling (STM) on our post text corpus. Table \ref{tab:topic} lists the ten selected topics, which ranged from cannabis ingredients to policies and the legalization of cannabis. Similar to the image labeling process, we addressed a multi-labeling problem for post texts, as one post text could pertain to zero, one, or multiple topics. For convenience, the BART API from Huggingface was employed to create scores for each topic-text pair. These BART scores were directly utilized in the downstream analysis as ten numerical features of each image (after mapping post texts to images). Figure \ref{fig:topic_score} (Appendix \ref{app:topic_score}) illustrates the distribution of the scores across the ten topics in our data, indicating that some topics appeared much more frequently than others.

\begin{table}
\begin{footnotesize}
  \centering
  \begin{tabular}{@{}l@{}@{}r@{}}
    \toprule
    1. Cannabis Ingredients     \\ 2. Policies and Legalization of Cannabis     \\
    3. Positive feelings of cannabis     \\ 4. Cannabis and Social Justice\\
    5. Cannabis Trade and Economics     \\ 6. Cannabis Prevention Campaign      \\
    7. Edibles and Cannabis     \\ 8. Health Risks Associated with Cannabis \\    
    9. Cannabis for Pain Management \\      10. Mental Health and Cannabis \\    
    \bottomrule
  \end{tabular}
\end{footnotesize}  
  \caption{STM-generated Topics as Covariates.}
  \label{tab:topic}
  
\end{table}

\section{Analyze association with online activities}

This study focused on the association between different visual elements in the social media environment and online interactions in response to the corresponding social media post. Specifically, we performed Linear Regression to address this goal, using users' online interaction as the response. The following sections discuss how we created an \textit{engagement score} for each post to quantify its related online interactions. 


\subsection{Quantification of Online Interactions.} 


We measured the volume of the online activities following an image post by normalizing the volume with the average amount of activities incited by the corresponding Facebook account. This normalization is performed to address the large variation in the popularity of different accounts. Specifically, for each image, we identify the account which posted it. We then take the last 100 image posts from that account and calculate the average number of interactions at each age (15 minutes old, 60 minutes old, 5 hours old, etc.). Next, we compared the pattern of interactions following the image post we are interested in that average and multiplied the difference by the weights in each dashboard. This comparison generated an \textit{engagement score} for the image post. A score over 1 means the post is performing better than average from that account, and at that point in time (10 minutes after posting, 1 hour after posting, etc). A negative score means it is performing worse than the average.


\subsection{Preliminary Results}

The results (see Table \ref{tab:reg_binary}) fitted a linear regression model to predict engagement score based on image and text features. These findings revealed significant positive associations between online activities and the presence of visual cues such as \textit{candy} and \textit{weed} visual cues. A negative association was observed for \textit{text} within images. Notably, the colorfulness of post images was significantly negatively associated with online engagement. Regarding textual topics, \textit{policies and legalization of cannabis}, \textit{cannabis and social justice}, and \textit{positive feelings about cannabis} were significantly positively associated with online activities. In contrast, topics such as \textit{cannabis for pain management} and \textit{mental health and cannabis} demonstrated significant negative associations with online engagement.

\begin{table}
\begin{scriptsize}
\centering
\begin{tabular}{@{}lccc@{}}
\midrule
                                                                         & \textbf{estimation} & \textbf{P$> |$t$|$} & \textbf{\%95 CI}\\
\midrule
\textbf{Intercept}                                                       &     -10.1942    &         \textbf{0.000}        &      (-11.829, -8.560)\\
\midrule
\textbf{type (reference=photo)} & & &\\
link                   &      -2.3146    &         \textbf{0.000}        &       (-3.065, -1.564)\\
live video complete  &       7.8269    &         0.077        &       (-0.858, 16.512)\\
live video scheduled &       3.3386    &         0.397        &       (-4.390, 11.067)\\
native video          &      16.7042    &         \textbf{0.000}        &       (12.114, 21.295)\\
status                 &      -0.1534    &         0.909        &       (-2.790, 2.483)\\
video                  &      -3.4102    &         \textbf{0.040}        &       (-6.670, -0.150)\\
youtube                &       3.1502    &         \textbf{0.027}        &        (0.354, 5.947)\\
\midrule
\textbf{diurnal cycle (reference=Morning)} &  & &\\
Afternoon  &       2.4860    &         \textbf{0.000}        &        (1.571, 3.401)\\
Evening    &       1.3554    &         \textbf{0.008}        &        (0.357, 2.354)\\
Night      &      -0.5146    &         0.340        &       (-1.572, 0.542)\\
\midrule
\textbf{colorfulness score}                                             &      -0.0336    &         \textbf{0.000}        &       (-0.046, -0.021)\\
\midrule
\textbf{brightness score}                                               &      -0.0065    &         0.074        &       (-0.014, 0.001)\\
\midrule
\textbf{is weekend}                                                     &      -0.7678    &         0.069        &       (-1.595, 0.060)\\
\midrule
\textbf{Post image labels} & & &\\
candy                                                    &       5.8403    &         \textbf{0.017}        &        (1.047, 10.634)\\
bakery                                                  &      -0.8872    &         0.512        &       (-3.539, 1.765)\\
fruit                                                    &       2.5166    &         0.079        &       (-0.291, 5.324)\\
human                                                   &       0.4942    &         0.177        &       (-0.223, 1.212)\\
weed                                                     &       2.5978    &         \textbf{0.000}        &        (1.544, 3.652)\\
text                                                    &      -0.9217    &         \textbf{0.012}        &       (-1.643, -0.200)\\
\midrule
\textbf{Post text topic scores} & & &\\
Cannabis Ingredients                                 &       0.3542    &         0.683        &       (-1.348, 2.057)\\
Positive feelings of cannabis                       &       3.3647    &         \textbf{0.000}        &        (1.860, 4.869)\\
Cannabis Trade and Economics                       &      -0.9708    &         0.471        &       (-3.613, 1.671)\\
Edibles and Cannabis                                 &      -1.5927    &         0.071        &       (-3.323, 0.138)\\
Cannabis for Pain Management                        &      -4.8443    &         \textbf{0.000}        &       (-7.219, -2.470)\\
Policies and Legalization of Cannabis              &       3.1464    &         \textbf{0.002}        &        (1.176, 5.117)\\
Cannabis and Social Justice                         &       2.5947    &         \textbf{0.019}        &        (0.418, 4.771)\\
Cannabis Prevention Campaign                         &       1.8193    &         0.198        &       (-0.949, 4.587)\\
Health Risks Associated with Cannabis              &       3.2779    &         \textbf{0.003}        &        (1.144, 5.412)\\
Mental Health and Cannabis                          &      -3.9959    &         \textbf{0.000}        &       (-5.802, -2.190)\\
\bottomrule
\end{tabular}  
\end{scriptsize}
\caption{Linear Regression Results. P-values that are smaller than 0.05 have been marked with boldface.}\label{tab:reg_binary}  
\end{table}
\section{Discussion and Conclusion}
 This study highlights key associations between visual cues in social media posts and user engagement. Specifically, we identified a positive correlation between images featuring candy and weed visuals and online interactions, while text-heavy images showed a negative association. Interestingly, the colorfulness of images was negatively associated with user engagement, contrary to findings in other visual marketing contexts. In terms of textual content, topics related to policies, social justice, and positive sentiments about cannabis were positively correlated with engagement, whereas topics concerning pain management and mental health showed negative relationships.
These findings have important implications for regulating cannabis marketing, particularly in ensuring that content does not unintentionally appeal to youth or vulnerable populations. Current regulations in the United States fall short in effectively restricting youth-targeted content in the marketing of cannabis-infused edibles. To address this, it is imperative to establish stricter and more uniform national guidelines that mandate responsible packaging and marketing practices to minimize the risk of youth initiation or accidental consumption. Policymakers can leverage these findings to develop more informed and effective regulatory frameworks.\\ Despite its contributions, this study has limitations that warrant further investigation. First, the analysis was limited to Facebook posts over a six-month period in 2021, potentially excluding evolving trends and patterns across other platforms and timeframes. Future studies could expand the scope to include longitudinal data and diverse social media platforms, such as Instagram or TikTok, where visual content plays a central role.
Second, while this study focused on specific visual and textual features, it did not account for audience-specific factors such as demographic or psychographic differences, which could mediate the observed associations. Future research could incorporate audience segmentation to understand how different groups engage with cannabis-related content. Future research could explore these downstream effects, providing a more comprehensive understanding of how online cannabis marketing influences public health outcomes.\\
Overall, this study underscores the complex dynamics of online cannabis marketing, highlighting the interplay between visual and textual content in shaping user engagement. The findings offer valuable guidance for policymakers and researchers alike in addressing the challenges posed by cannabis marketing, fostering a safer and more responsible digital landscape.

{
    \small
    \bibliographystyle{ieeenat_fullname}
    \bibliography{main}
}
\vspace{1cm}

\clearpage
\setcounter{page}{1}
\maketitlesupplementary

\section*{Appendix A}
\label{sec:app}

\subsubsection*{Search Keyword} kief OR mmj OR mmot OR weed brownie OR weedbrownie OR weedbrownies OR Blunt OR budder OR Burning one down OR Butane honey oil OR Cannabis OR cannabis butter OR cannabis candy OR cannabis edible OR cannabis infused OR cannabutter OR Cbd OR Cheeching OR Chillum OR Concentrates OR Cosmic brownie OR Dab OR dab life OR dabber OR dabbers OR dabbin OR Dablife OR dabs OR dank cookie OR dank edible OR edibles OR enail OR e-nail OR ganja OR hash butter OR hash cookie OR hash oil OR hemp OR high edible OR indica OR infused marijuana OR kush OR kush butter OR kush cookies OR kush infused OR Legalize marijuana OR legalized marijuana OR legalizing marijuana OR Live resin OR magic brownie OR magic brownies OR Marihuana OR Marijuana OR Marijuana brownie OR marijuana butter OR Marijuana cake OR Marijuana candy OR marijuana cookie OR marijuana edible OR Marijuana gummies OR medibles OR Medical cannabis

\section*{Appendix B}
\subsubsection*{Codebook}
\begin{itemize}
\item Candy (0=No  1=Yes) 
\end{itemize}
\begin{itemize}
\item Bakery (0=No  1=Yes) 
\end{itemize}
\begin{itemize}
\item Fruit (0=No  1=Yes) 
\end{itemize}
\begin{itemize}
\item Human (0=No  1=Yes) 
\end{itemize}
\begin{itemize}
\item Colorfulness (score)
\item Brightness (score)
\end{itemize}
\begin{itemize}
\item Text (0=No  1=Yes) 
\end{itemize}

\section*{Appendix C}
\subsection{Details of Using the CLIP}
We used the resnet-50x64 model of the CLIP. 
The embeddings of each of the seven queries are calculated by the CLIP model, and the average of the seven embeddings are used when calculating the similarities with image embeddings.

\texttt{['itap of a {}.', 'a bad photo of the {}.', 'a origami {}.', 'a photo of the large {}.', 'a {} in a video game.', 'art of the {}.', 'a photo of the small {}.']}\label{app:query}

We added \textit{dog} to our visual element list when constructing queries. This is because, empirically, we found that there are a decent amount of images with dogs in our image data and the CLIP model tends to give these images high similarity scores w.r.t. the human label. While applying the CLIP model, we added the dog label to bring down the calculated probabilities for human features, so as to prevent human features from being erroneously detected in images that have dog (but no human).

\subsection{Distribution of topic scores for post texts}\label{app:topic_score}

\begin{figure}[h]
  \includegraphics[width=\linewidth]{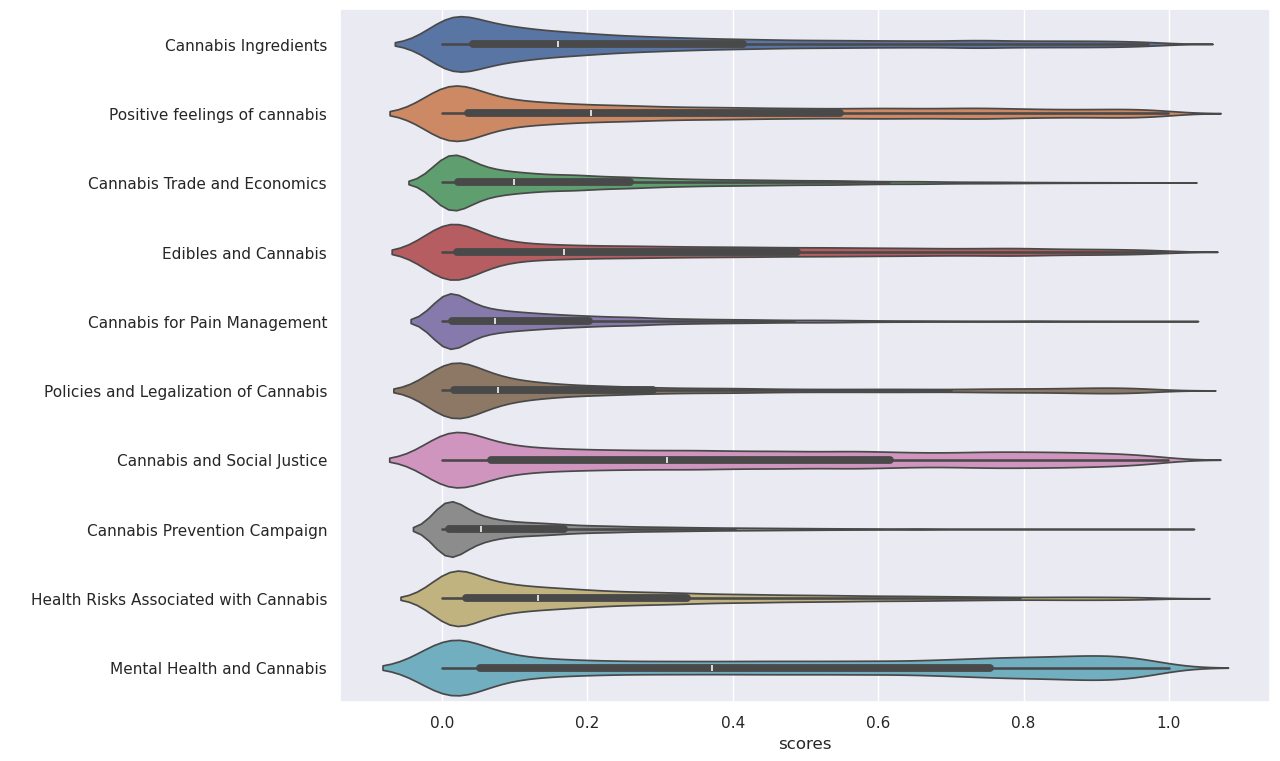}
  \caption{Distribution of topic scores for post texts as assigned by applying the BART model.}
  \label{fig:topic_score}
\end{figure}

Figure \ref{fig:topic_score} shows distribution of post text topic scores.

\end{document}